\documentclass[letterpaper,twocolumn,10pt]{article}
\usepackage{usenix2019_v3}
\usepackage{tikz}
\usepackage{amsmath}
\usepackage{filecontents}
\usepackage{tabularx}
\usepackage{lscape}
\usepackage{soul}
\usepackage{float}
\usepackage[table,xcdraw]{}

\begin{document}
%-------------------------------------------------------------------------------

\date{}

\title{\Large \bf PCaaD: Towards Automated Determination and Exploitation of Industrial Processes}

\author{
{\rm *Benjamin Green, *William Knowles, **Marina Krotofil , *Richard Derbyshire, *Daniel Prince}\\
\rm *Neeraj Suri\\
\{b.green2, w.knowles, r.derbyshire1, d.prince, neeraj.suri\}@lancaster.ac.uk\\
marina.krotofil@tuhh.de\\
*School of Computing and Communications, Lancaster University, Lancaster, United Kingdom\\
**Hamburg University of Technology, Hamburg, Germany\\
}

\maketitle

%-------------------------------------------------------------------------------
\begin{abstract}
%-------------------------------------------------------------------------------
Over the last decade, Programmable Logic Controllers (PLCs) have been increasingly targeted by attackers to obtain control over industrial processes that support critical services. Such targeted attacks typically require detailed knowledge of system-specific attributes, including hardware configurations, adopted protocols, and PLC control-logic, i.e. process comprehension. The consensus from both academics and practitioners suggests stealthy process comprehension obtained from a PLC alone, to conduct targeted attacks, is impractical. In contrast, we assert that current PLC programming practices open the door to a new vulnerability class based on control-logic constructs. To support this, we propose the concept of Process Comprehension at a Distance (PCaaD), as a novel methodological and automatable approach for system-agnostic exploitation of PLC library functions, leading to the targeted exfiltration of operational data, manipulation of control-logic behavior, and establishment of covert command and control channels through unused memory. We validate PCaaD on widely used PLCs, by identification of practical attacks.
\end{abstract}

%-------------------------------------------------------------------------------
% Papers Core Content
%-------------------------------------------------------------------------------
\vspace{-10pt}
\section{Introduction}
\label{introduciton}
\vspace{-10pt}
Acting as the bridge between physical industrial processes and enterprise systems, Industrial Control Systems (ICSs) deliver wide-spread monitoring, control, and automation capabilities to a broad spectrum of end-users. The Purdue Enterprise Reference Architecture model (PERA)~\cite{Williams1994} provides an approach to compartmentalize the complexity of ICSs into hierarchical layers. Each layer affords system users with access to industrial processes and the data they generate. The lower the layer, the closer associated devices are to the processes they oversee, with Programmable Logic Controllers (PLCs) providing a primary interface to operational components (e.g. pumps and valves) via sensors and actuators.

A number of historical attacks have demonstrated the willingness of attackers to target ICSs~\cite{Derbyshire2018}, with initial access obtained via malicious USBs, project files, software updates, the supply chain, public facing systems, etc.~\cite{ICS-CERT-Havex, Mirian:scanning:2016, slay2007lessons, Falliere2011} However, there still exists a primary challenge, once access is obtained, how can a cyber-physical attack be undertaken i.e., an attack in which industrial operational process manipulation is achieved. A comprehensive body of existing work details the challenge attackers face in the development of cyber-physical attacks, captured under the umbrella term of "process comprehension". This is defined as "the understanding of system characteristics and components responsible for the safe delivery of service"~\cite{Green2017a}. The described challenges align to a lack of a single resource by which attackers can obtain sufficient process comprehension to conduct a cyber-physical attack. We see this not only in the understanding of physical operational processes, but also the interconnectivity and broader configuration parameters of devices an attacker may choose to target.

While there exist a number of tools and techniques one can use to develop a level of process comprehension through the targeting of PLCs alone, they are limited by functionality, scope, and detectability~\cite{NMap, beresford2011, ICSSploit2020}. The holy grail would be to stealthily (avoiding detection) obtain complete Process Comprehension at a Distance (PCaaD) targeting only PLCs, while simultaneously preventing any disruption to their operation. We assert that current PLC programming practices provide a segway into capability of this kind, and provide validation through the exploration of widely used control-logic (PLC code) library functions, developed by device vendors for use by programmers. This leads to the following five exploitation capabilities: (1) the remote enumeration of control-logic library functions, (2) the exfiltration of operational process data and configuration parameters, (3) the targeted manipulation of control-logic behavior, impacting operational processes and configuration parameters, (4) the establishment of covert command and control (C2) channels through unused memory, and (5) the end-to-end system-agnostic automation of 1-4.

This paper serves as an significant step in developing PCaaD capability, forming a greater understanding of the role PLC programming practices play in process comprehension techniques, using library functions as an explorative base. Through this, we begin to develop capability aligned to automated environment-agnostic cyber-physical attacks, and build upon an emerging vulnerability class based on control-logic constructs. Therefore, the novel contributions of this work are:

\vspace{-5pt}
\begin{itemize}
	\itemsep0em
	\item A stealthy method to enumerate library functions based on memory allocation.
	\item A targeted approach to data exfiltration and operational process/configuration manipulation.
	\item A method allowing for the establishment of a covert C2 channel via unused memory.
	\item An automated process to enact remote enumeration, exfiltration, exploitation, and covert C2 channel creation.
\end{itemize}
\vspace{-5pt}

The remainder of this paper is structured as follows. Section~\ref{relatedwork} covers related work. Section~\ref{Threatmodel} details a threat model/set of attack vectors. Section~\ref{PLCbackground} provides a background on PLC program structures. Section~\ref{PCaaD} develops our main contribution of PCaaD, which is subsequently validated in Section~\ref{PoC}. Section~\ref{discussion} provides a set of lessons learnt, including a process flow for automated PCaaD and attack execution. Section~\ref{conclusion} concludes the paper and offers areas for future work.
\vspace{-10pt}
\section{Related Work}
\label{relatedwork}
\vspace{-10pt}
Over the last decade, there has been an increasing volume of research targeting the exploitability of embedded systems used in industrial settings. This reflects both the large number of "low-hanging fruit" vulnerabilities, and an increased interest from attackers towards the disruption of industrial processes. To date, research efforts have predominately focused on real-time operating systems, firmware vulnerabilities, industrial protocols, and bypassing traditional security controls~\cite{abbasi2016ghost, Nochvay2019, drias2015taxonomy, wardak2016plc, Biham2019}. 

Only a small subset of existing work focuses on controller programming security implications. Kottler et al.,~\cite{kottler2017formal} explore the formal verification of ladder logic (a control-logic programming language). Eckhart et al.,~\cite{eckhart2019security} consider security implications within the wider system development lifecycle. While Serhane et al~\cite{serhane2018plc,serhane2019programmable} examine coding practices that could cause unsafe conditions, in the majority of discussed practices, an attacker is required to push new control-logic to target systems. In a similar vein, the development of malicious control-logic to cause denial-of-service conditions has also been explored~\cite{govil2017ladder}. More recently, Fluchs~\cite{Fluchs2020} describes an initiative backed by the International Society of Automation to define "The Top 20 Secure PLC Programming Practices", with a community driven approach to identify additional practices moving forwards. Finally, the work of Ljungkrantz and Akesson~\cite{Ljungkrantz2007} provides an empirical exploration of PLC programming practices using library components. This work showcases the wide spread adoption of libraries and the potential impact of homogeneity in control-logic design. However, it does not consider the cyber security implications of such practices.

Few works have addressed the need for process comprehension from an attackers perspective; a critical pre-condition when seeking to achieve operational impact beyond simple denial-of-service~\cite{Green2017a, gollmann2015}. Research here has focused on the exploitation of configurational practices~\cite{wardak2016plc}, or wider attack scenarios and taxonomies~\cite{drias2015taxonomy}.

Research exploring physics-aware attack payloads for industrial processes are also limited~\cite{McLaughlin2012, Garcia2017}. While some elements of control-logic analysis in these works is done automatically, payload design still relies on a "human-in-the-loop". In the closest work to ours~\cite{Otorio2019}, the authors created an approach to automate the analysis of control-logic and Human Machine Interface (HMI) project files, before building an attack payload. However, the prerequisite in obtaining such files presents an obstacle.

Currently, Stuxnet and Industroyer form the most sophisticated ICS-focused attacks to date~\cite{Falliere2011, Cherepanov2017}. Stuxnet applied a novel approach to target identification using known characteristics within system data blocks (SDBs), a component of Siemens PLC control-logic. However, this was highly targeted as SDBs are unique to each implementation only providing PLC hardware parameters. While Stuxnet embodied precision, Industroyer manipulated every identified variable on the Remote Terminal/Telemetry Unit (RTU) (set all states to 0), without understanding the targets associated operational processes.

To summarize, the security implications of PLC programming has received limited attention, a critical gap noted by others~\cite{eckhart2019}. Where it exists, there has been no examination of how deployed control-logic could be stealthily enumerated to support process comprehension. We assert current programming practices play a key role in a PLCs exploitability, providing PCaaD without a priori target system knowledge. The following sections describe a threat model and PLC programming structures, prior to the identification of PLC-centric artifacts that are used to build sufficient process comprehension to execute a targeted attack.

\vspace{-10pt}
\section{Threat Model}
\label{Threatmodel}
\vspace{-10pt}
To support discussions throughout the remainder of this paper, the following system under consideration and set of example attack vectors are presented. This offers insight into how PLCs can be targeted by multiple threat actor categories, with varying capabilities and resources~\cite{cost}.

\vspace{-10pt}
\subsection{System Under Consideration}
\vspace{-5pt}
Figure~\ref{fig:scenario} provides an overview of infrastructure architecture frequently found in distributed ICS applications, such as water and energy~\cite{Stouffer2015}. Within the Field Site (e.g. a water pumping station) there is a Windows-based HMI~\cite{Siemenswincc}, used by trusted operators to monitor and control physical operational processes via the PLCs~\cite{Siemens2020a}. There are two PLCs which are used to monitor, control, and automate operational processes. There is also an additional PLC, the PLC/RTU, which performs a similar process automation role but also communicates with a remote Top End System (TES), the Supervisory Control and Data Acquisition (SCADA) Server~\cite{zenon}. Historically, PLCs only communicated with devices inside the Field Site, with dedicated RTUs forwarding operational data to TESs. However, due to the increased computational resource and connectivity available in modern PLCs, they now act in a dual-purpose capacity, providing RTU capability/interconnectivity with TESs~\cite{Gouglidis2018}. There is a network switch, and WiFi router, providing the Field Site with local and remote communications. The remote SCADA Server communicates with the PLC/RTU via its boundary router (in a real-world setting there would be multiple Field Sites communicating with a TES for infrastructure-wide visibility). Finally, the Windows-based Alarm Management Workstation accesses operational data/systems via the SCADA Server.

\vspace{-10pt}
\subsection{Attack Vectors}
\vspace{-5pt}
Overlaid onto Figure~\ref{fig:scenario} we have four example attack vectors (AV1-4), each with a set of threat actor and defence profiles.

\vspace{-10pt}
\subsubsection{Attack Vector 1}
\vspace{-5pt}
Despite growing awareness of cyber security threats to ICSs, ICS devices are being exposed to the Internet without suitable security measures~\cite{Shodan2020, Mirian:scanning:2016}.

For this attack vector, we assume the PLC/RTU has been configured on a public IP address for remote SCADA Server access. This allows the threat actor to directly access the PLC/RTU and execute malicious commands with no defensive controls to circumvent. This could be enacted by a low-skilled threat actors.

\vspace{-10pt}
\subsubsection{Attack Vector 2}
\vspace{-5pt}
For added convenience, the use of wireless technologies is becoming more prevalent in ICSs. Conventional WiFi (802.11) for example, can be used to established connectivity between engineering laptops, mobile HMIs, PLCs, etc~\cite{Siemenswifi}. However if incorrectly configured (i.e. security and transmission range), they can induce additional risk outside of a Field Site's physical perimeter~\cite{slay2007lessons}.

For this attack vector, we assume the WiFi access provided by the router is open and insecure. This allows the threat actor to directly access the internal Field Site network and execute malicious commands with no defensive controls to circumvent. This could be enacted by a low-skilled threat actor.

\vspace{-10pt}
\subsubsection{Attack Vector 3}
\vspace{-5pt}
While Field Sites can be isolated using network-based controls, access can still be obtained through physical means~\cite{Falliere2011}. For example, trusted system operators, engineers, via the supply chain (e.g. 3rd party service providers and pre-infected device introduction), etc.~\cite{ICS-CERT-Havex, fsecure2014havex, slay2007lessons}

For this attack vector, we assume the connection between the Field Site and Data-Centre is secured based on the use of a VPN and an associated IP address/port rule-set. The rule-set permits the remote SCADA Server to communicate with the PLC/RTU, all other traffic is blocked. In addition, the WiFi access has also been secured with a strong WPA2 key. Here a trusted HMI operator inserts a USB stick containing malicious code into the HMI, which then executes malicious commands autonomously against all devices within the Field Site network via the switch. This could be enacted through the use of a malicious insider, or alternatively a high-skilled threat actor who is able to infect a trusted users USB stick (e.g. via the supply chain).

\vspace{-10pt}
\subsubsection{Attack Vector 4}
\vspace{-5pt}
Direct access to Field Site devices via networked communications may only be possible through existing trusted systems. Through the initial compromise of trusted systems, and subsequent lateral movement, the desired level of access can be achieved. Furthermore, social engineering is often viewed as a primary initial access technique, and impacts ICS environments in the same way as conventional IT systems.~\cite{Lee2014GERMANSTEEL, Liang2017}

For this attack vector, we assume the connection between the Field Site and Data-Centre is secured based on the use of a VPN and an associated IP address/port rule-set. The rule-set permits the remote SCADA Server to communicate with the PLC/RTU, all other traffic is blocked. In addition, the WiFi access has also been secured with a strong WPA2 key. Here the threat actor compromises the Internet connected Alarm Management Workstation via a malicious email. From this initial access, the threat actor then compromises the SCADA Server, which is used to execute malicious commands against the PLC/RTU.

\begin{figure}[h!]
    \centering
    \includegraphics[width=\linewidth]{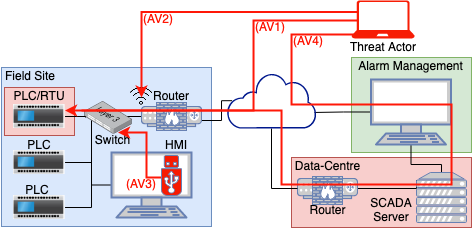}
    \vspace{-2\baselineskip}
    \caption{Threat Model}
	\label{fig:scenario}
	\vspace{-10pt}
\end{figure}
\vspace{-10pt}
\section{PLC Program Structures}
\label{PLCbackground}
\vspace{-10pt}
PLCs are available from a range of vendors, with varying deployments in multiple industrial settings dependent upon operational requirements~\cite{Stouffer2015}. They sequentially execute a series of instructions, referred to as a "Program". However, this paper uses the term \textit{control-logic} to provide a clear distinction during discussion. At a fundamental level, control-logic interfaces with Input/Output (I/O) channels, and based on input states, adjusts output states. Control-logic can provide additional, more complex functionality. This includes establishing configuration parameters for specific network protocols, emailing system's users in the event of an operational incident, and connect to remote engineer workstations in the event of a system failure.

The BSI/IEC standard 61131-3:2013~\cite{BritishStandardsInstitute2013} outlines five PLC programming languages. These are split into two categories, \textit{Graphical} (Ladder Diagram, Function Block Diagram, and Sequential Function Chart), and \textit{Text Based} (Instruction List, Structured Text). These languages are vendor and application agnostic, although vendor specific language subsets are often provided.

BSI/IEC 61131-3:2013~\cite{BritishStandardsInstitute2013} also defines the concept of Program Organization Units (POUs). The following definitions use, and expand, the terminology of BSI/IEC 61131-3 to provide a generalized model, abstracting away from vendor specific terminology, to support subsequent discussions:

\vspace{-5pt}
\begin{itemize}
	\itemsep0em
	\item \textbf{Programs}: Are the highest level of organizational unit. They control program execution enabling responses to cyclic, time-based, or interrupt-driven events during program execution.  They are composed of specific instructions but also Function Blocks and Functions.
	\item \textbf{Function Blocks (FB)}: Contain code that store their values permanently in memory, remaining available post Function Block execution.
	\item \textbf{Functions}: Provide discrete common functionality, for example, ADD or SQRT. Function POUs can use global variables to permanently store data, but do not have their own dedicated memory (i.e. local variables).
	\item \textbf{Variable Blocks (VB)}: Store program data and can be global (gVB), or local (fVB). The latter of which are associated with Function Blocks to provide long term data storage. The VB is an addition to the POU model as defined by IEC 61131-3. This standard describes the use of variables in a general sense, with limited ties to their storage.
\end{itemize}
\vspace{-5pt}

Control-logic can be written in Program, FB, and Function POUs. Typically, there are additional specialist elements for accessing and addressing other system's components, including peripherals and timers. However, comprehension of these is not necessary for the current discussion, and are therefore omitted.

A further noteworthy control-logic attribute is the use of pointers in, and between, VBs. This is useful for common information, such as configuration data, or for central recording of operational parameters. For example, consider two FBs. The first processes a water level, translating an I/O channels raw analogue reading into a total water volume. The second is required to access the total water volume for an additional calculation. Instead of writing the total water volume value to the second FBs fVB (copying/duplicating the value), the fVB would contain a pointer to a memory location where the value has been stored by the first FB.

The use of Functions and FBs support code reuse patterns within an organization across multiple deployments~\cite{Jacinto2017}. However, vendors often supply a library of Functions and FBs spanning commonly required functionality, aiding the development of control-logic. These libraries are also referred to as "instruction sets". The results of a study into two automotive assembly facilities, identified the repetitive use of libraries across their controller base~\cite{Ljungkrantz2007}. This process allows for the managed development/deployment of control-logic, ensuring suitability in one operational zone prior to widespread replication across the remaining estate.

It is worth noting a key difference compared to library Functions in conventional IT software. On the surface, these POUs appear similar to their IT equivalents, providing code reuse. However, they execute sequentially in the control-logic, rather than undertaking complicated execution stack management and sub-routine calls. The implication here is that when a Function or FB is used multiple times, it must be copied into the control-logic multiple times. With FBs a repeated fVB is allocated each time. In this way library Functions are more like tested and verified code {\it snipets}, cut and pasted into the control-logic to save time. Their deployment in any given infrastructure (e.g. real-world environments such as water and energy, or a testbed) is identical. This represents a key concept we exploit with our PCaaD approach.

\vspace{-10pt}
\section{The PCaaD Approach}
\label{PCaaD}
\vspace{-10pt}
Given the previously described generic operating and programming model applied to PLCs, there exists the possibility of enumerating control-logic by observing VB memory. Where a PLC permits remote access to its memory across the network (e.g. HMI and TES interactions to monitor and control operational processes are done in this way), an avenue is provided for remote extraction of its content. Furthermore, PLCs often include additional network functions allowing for remote interrogation, these have been used in Nmap fingerprinting scripts~\cite{NMap}.

The memory layout of fVBs is consistent across implementations. Through the identification of patterns in the memory layout of a fVB, it becomes possible to identify (enumerate) them and their associated FBs. Once a FB is known, use of the data contained within its fVB can then be interpreted and exploited. Our advocated PCaaD process consists of a two phased approach to enumeration, \textit{Data Retrieval} and \textit{VB Determination}, the completion of which allows for targeted \textit{Exploitation}.

\vspace{-10pt}
\subsection{Enumeration Phase 1: Data Retrieval}
\label{DataRetrieval}
\vspace{-5pt}
Data Retrieval is the first phase of PCaaD, and focuses on retrieving only the necessary information required for subsequent \textit{VB Determination}. This can be applied to each of the four attack vectors described in Section~\ref{Threatmodel}, where we state "execute malicious commands".

The data retrieval phase is common in many attack approaches and is often found in wider reconnaissance activities~\cite{icskillchain}. Typical reconnaissance techniques focus on identifying services running on a given system, and any additional freely available information which may be useful to the attacker. Current PLC reconnaissance tooling is limited, identifying basic parameters such as manufacturer, model, and firmware version~\cite{NMap, Efanov2017}. 

Through the exploration of a range of PLCs, this work identifies the following three common data retrieval methods which may be used during this stage. Each has implementation specific pros and cons, which are highly dependent on the attackers objectives and modus operandi (See Section~\ref{PoC}).

\vspace{-5pt}
\begin{itemize}
	\itemsep0em
	\item \textbf{Metadata:} The majority of vendors provide network functions to query control-logic meta information. These functions do not provide information regarding the current operation of control-logic, but rather information about how it (and the wider PLC) is configured. Within a traditional IT context, this is comparable to querying the manifest of a shared code object (e.g. DLL on Windows or Shared Objects in Linux).
	\item \textbf{Bulk Transfer:} A PLC operating system will often provide a bulk transfer operation, allowing engineers to extract the current state of POUs (gVBs, fVBs, Program, FB, and Functions), supporting diagnostic fault finding, scheduled backups, etc. Within a traditional IT context, this is comparable to a web-server with direct file store access.
	\item \textbf{Memory Address Interrogation:} PLCs provide the ability to remotely interrogate internal memory locations for their current state. This functionality is used to provide operational monitoring and control capability, supporting the retrieval of one or more data items through the specification of memory locations. Within a traditional IT context, this is comparable to SNMP Object requests using a known Management Information Base.
\end{itemize}

\vspace{-10pt}
\subsection{Enumeration Phase 2: VB Determination}
\label{vbcomp}
\vspace{-5pt}
VB Determination, is the process of identifying which fVBs and associated FBs, have been included within a PLCs control-logic, through the analysis of retrieved data from Phase 1. Note that Functions do not have associated VBs and so cannot be identified in this way.

A simple first order approach can be derived from the use of \emph{Metadata} retrieval approaches. As with interrogating shared code object manifests, a fVBs \emph{Metadata} contains attributes one can use in the determination of its associated FB.

Through the use of a \emph{Bulk Transfer}, or a byte wise download, we can obtain fVBs in their entirety. Once obtained, a search for unique attributes (similar to those in \emph{Metadata}), can be conducted to determine its associated FB.

As previously discussed, each fVB has a static memory layout. This contains variables used by the associated FB, and is consistent across all operational and deployment contexts (e.g. water, energy, a testbed). A consistent static memory layout allows for the identification of distinct characteristics forming signatures, to identify fVBs and their corresponding FB using \emph{Memory Address Interrogation}. This concept can be considered similar to rainbow tables, providing a set of pre-computed signatures ready for use during an attack. The following characteristics have been initially selected for fVB signatures:

\begin{itemize}
	\itemsep0em
	\item \textbf{fVB Size:} The fVB size (quantity of allocated bytes) is fixed for all instantiations of the FB across the control-logic base, and is dependent on the number and type of variables used.
	\item \textbf{Known Values:} Some FBs are pre-set with default (and thus known) values for variables in the associated fVB. Here it would be possible to map these known defaults and use them as an indicator of potential fVB match.
	\item \textbf{Variable Usage:} Examination through memory interrogation, to reveal potential data types. For example, if a memory location was only ever set to 0x00 or 0x01, it would indicate a potential boolean data type. Once the data type and memory offset is determined for enough data types, it is possible to map this to a known fVB layout.
	\item \textbf{Data Type Features:} In some PLC hardware architectures it is known that variable allocation is based on defined bit boundaries. Consider an architecture allocation applying a 16 bit boundary. A boolean data type would occupy more than a single bit. The use of numerous boolean variables in the fVB would create a signature of null space which can be used to identify the fVB.
\end{itemize}

It is worth noting, that for in-house developed FBs (not publicly available library FBs), an attacker would not have an established signature for their associated fVB ahead of an attack - except with the use of other intelligence (i.e. precursory attacks to obtain PLC source code from either the target facility, or a subcontractor where PLC programming is outsourced). However, given the identification of an unmatched VB, it is then possible to generate a signature. When applying PCaaD to other devices within the target infrastructure, this new signature can then be used to identify where the unknown FB is reused.

\vspace{-10pt}
\subsection{Exploitation based on PCaaD}
\vspace{-5pt}
Using the aforementioned techniques, an attacker is able to identify which library FBs are included within the broader control-logic base. With this newly acquired information, an attacker is now presented with a range of options to launch an attack. We present three attacks which can be executed, one of which is a storage-based covert C2 channel.

\vspace{-10pt}
\subsubsection{Attack 1: Exfiltrate FB Variables}
\vspace{-5pt}
\textit{This attack extracts data tied to operational process state, and/or PLC configuration, dependant upon the FBs in use. It can be applied to each of the four attack vectors described in Section~\ref{Threatmodel}, where we state "execute malicious commands".}

It has been asserted that library FBs require well defined fVBs, containing characteristics one can use to develop signatures. Once a FB is known, the associated fVB can be targeted to extract variable states, using a small number of \textit{Memory Address Interrogation} requests. 

As discussed in Section~\ref{PLCbackground}, VBs may contain pointers to alternate memory locations. Pointers typically have well defined static structures, which can be decoded. Therefore, if a FB is using pointers in its associated fVB, pointing to variables in a gVB, their state can also be retrieved, albeit with an additional \textit{Memory Address Interrogation}. Furthermore, the use of pointers supports process comprehension, as it allows an attacker to identify gVBs that are used by FBs.

\vspace{-10pt}
\subsubsection{Attack 2: Targeted Manipulation of FB Operation}
\vspace{-5pt}
\textit{This attack gains fine grained control of FBs, to subvert PLC or operational process behaviors. It can be applied to each of the four attack vectors described in Section~\ref{Threatmodel}, where we state "execute malicious commands".}

Previously demonstrated attacks either required a priori information on the target PLC~\cite{Falliere2011}, adopt brute-force techniques~\cite{Robles-Durazno2019}, or focus on denial of service (DoS) impact~\cite{beresford2011}. For example, in the case of Stuxnet~\cite{Falliere2011} it was widely reported the only way in which this attack was achievable, was through a complete attacker implementation of the target infrastructure based on significant intelligence. 

The approach discussed thus far enables PCaaD against an unknown system (no requirement for a priori intelligence or replication of the target infrastructure). An attacker knows how the FB variables are being used, therefore has a greater level of understanding on how they can be manipulated. For example, consider a FB responsible for counting how many litres of water have been treated. To set this value back to 0, the attacker has two options, overwrite the integer representing the total value, or toggle the count reset bit to 1 and then 0.

\vspace{-10pt}
\subsubsection{Attack 3: A Novel Storage Based Covert Channel}
\vspace{-5pt}
\textit{This attack utilizes unused PLC memory (null space), to create a covert channel. It can be applied as a combination of attack vectors 3 and 4 described in Section~\ref{Threatmodel}, where we state "execute malicious commands", to create a covert channel between the SCADA Server and the HMI}

As previously discussed, the presence of null space acts as a key characteristic in the identification of fVBs and their corresponding FBs. However, this null space can also be used to establish a covert communications channel. Null space is allocated due to alignment with bit boundaries, and is often not considered when processing data in the memory location. For example, an 8 bit value allocated in a 16 bit boundary system, would result 8 bits of unused space. This is a concept similar to the use of file slack space for hiding data, and is used in tools such as bmap~\cite{mulazzani2013}.

This approach to covert channel creation, exploits the observation that a PLC is in a trusted position within an operational network. As shown in our scenario (Figure~\ref{fig:scenario}), the PLC/RTU is required to communicate within its local Field Site network (i.e. with the HMI and other PLCs), as well as the data-centre network (i.e. SCADA Server). This means an external party could pivot via the PLC/RTU from the Data-Centre network to the Field Site network. A covert channel of this nature could be used for two primary purposes: \textit{C2} and \textit{Data Exfiltration}.

For the C2 channel, an attacker uses the PLC/RTUs null space to issue commands from the C2 server (SCADA Server) to the C2 client (HMI). The C2 client periodically checks the null space for these commands and execute them accordingly. A channel built on this approach needs to consider the following:

\begin{itemize}
	\itemsep0em
	\item How the C2 Sever (SCADA Server) and Client (HMI) synchronize on a subset of the null space.
	\item What periodicity for checking and writing to null space should be used, as this may be dynamic dependant on network conditions (e.g. round trip times between systems).
	\item How reliable does the channel scheme need to be, verses the communications overhead of introducing increased robustness. An increased communication overhead could lead to increased detectability.
	\item The possible commands, how they are encoded, and how results are relayed.
\end{itemize}

The final point listed here leads onto our second covert channel use: a data exfiltration channel. Once a C2 channel is established it also becomes possible to transfer more than simplistic commands and responses. In much the same way as FTP applications have a control and data channel, a separate channel could be used to bulk transfer larger volumes of data. Adopting the same approach as the C2 channel (null space), this secondary channel can be used to send and receive data. As with the C2 channel there are similar challenges on reliability, resiliency, and speed of communications, verses usability and detectability.

The concept of using null space and a PLC as a covert channel for attack, gives rise to PLCs being used as a means for attack, not just the target of an attack. This opens up a new class of security challenge which must be considered when deploying PLCs. Section~\ref{PoC} demonstrates how this approach, and the two prior attacks, can be practically achieved.

\vspace{-10pt}
\subsection{Summarizing PCaaD}
\vspace{-5pt}
This section has introduced the key PLC concepts required to understand the general security problem class of PCaaD. It has also demonstrated the feasibility of PCaaD by exploiting the code reuse patterns of common PLC software libraries. These software libraries provide a commonality across PLC implementations, regardless of operational or deployment context (e.g. water, energy, a testbed).

It is argued that this commonality provides a mechanism to identify FB signatures, which gives rise to a higher level of process comprehension. Memory features are shown to provide an approach by which FB signatures can be identified in a stable and repeatable way. 

It is anticipated that machine learning approaches could be used for more advanced fVB identification, with mapping based on the features identified here. In addition, it would be expected that other features of fVB will be identified and used to provide robust signatures. However, using the techniques outlined here, it is possible to perform enumeration of all FBs available from vendor libraries. 

Given the level of process comprehension which can be obtained through PCaaD, more sophisticated attacks can be performed. This includes configuration and operational data exfiltration, as well as fine grained variable manipulation. This section also introduced the concept of a storage based covert channel, via a PLCs unused memory (null space). This covert channel opens a new class of security challenge for PLCs, such that they are now not only the target of attack, but also the means by which an attack occurs.

Whilst this section outlined key, high-level concepts, the following section provides details on PCaaD's practical implementation using the widely deployed Siemens 300 Series PLC platform.

\vspace{-10pt}
\section{PCaaD Proof of Concept}
\label{PoC}
\vspace{-10pt}
To facilitate practical proof-of-concept exploration, we used a Siemens 300 series PLC~\footnote{Selected based on global adoption, making it a representative use-case~\cite{Siemens2020a}.}, and the Siemens TIA v13 platform as a programming agent~\cite{Siemens2020}. The library functions discussed herein are inbuilt into TIA v13 (Professional). We summarize this section with a note to the described techniques applicability to a broader PLC base, beyond the directly described Siemens 300 Series (e.g. Siemens 400, ET200, 1200, and 1500 Series, and ABB Drive Functions).

\vspace{-10pt}
\subsection{Siemens PLC Ecosystem}
\label{SiemensEco}
\vspace{-5pt}
Siemens 300 series PLCs support Ladder Diagrams, Statement List (Instruction List), Function Block Diagram, Graph, and Structured Text programming languages. When programming these devices four primary blocks are used to build control-logic: Organization Blocks (OB) (i.e. Program POUs), Function Blocks (FB), Functions (FC), and Data Blocks (DB) (i.e. Variable Block POUs). These are aligned to the previously described BSI/IEC 61131-3:2013~\cite{BritishStandardsInstitute2013}.

Within OBs, FBs, and FCs, one is able to write control-logic. DBs are used to store data, more specifically, variables called by OBs, FCs, and FBs. There exist a number of additional symbol types where data can be generated, outputted, and stored. These can be summarized as I/O Signals (I, Q, etc.), Marker Memory (M, MB, etc.), Peripheral I/O (PIW, PQB, etc.), and Timers and Counters (T \& C)~\cite{PLCDev2020}. 

\begin{figure}[h!]
    \includegraphics[width=\linewidth]{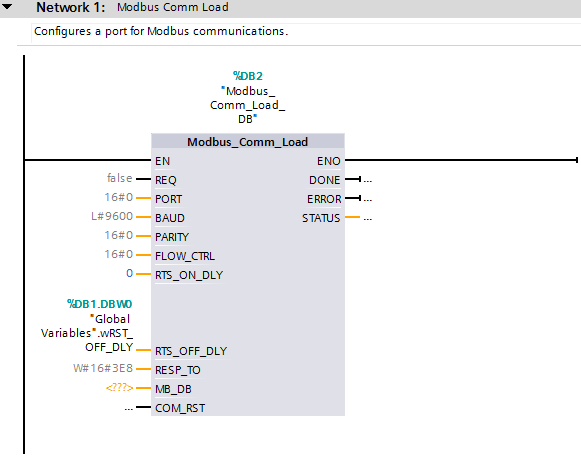}
    \vspace{-2\baselineskip}
    \caption{Modbus Library Function}
    \label{fig:modbusfunctionladder}
    \vspace{-5pt}
\end{figure}

Figure~\ref{fig:modbusfunctionladder} depicts the library FB \textit{\textbf{Modbus\_Comm\_Load}} residing on a Ladder Diagram rung. This FB is provided by Siemens as part of their TIA \textit{\textbf{Communications}} library, and is responsible for establishing the configuration of a port, from which the PLC can communicate over serial using the Modbus protocol. To the left of the \textit{\textbf{Modbus\_Comm\_Load}} FB are a set of inputs, these are configuration parameters (port, baud, parity, etc.). To the right of the \textit{\textbf{Modbus\_Comm\_Load}} FB are a set of outputs generated by the function (done, error, and status). By default, some of the inputs are pre-issued and can be seen in grey. These can be left unchanged if their states match the required configuration. Alternatively, inputs can be replaced directly within the rung, as can be seen with the blue \textit{\textbf{RTS\_ON\_DLY}} input (\textit{\textbf{0}}), or with variables stored in global DBs (gVBs), as can be seen with \textit{\textbf{DB1.DBW0}} (the address for global variable \textit{\textbf{wRST\_OFF\_DLY}}) applied to the \textit{\textbf{RST\_OFF\_DLY}} input.

\begin{figure}[h!]
    \includegraphics[width=\linewidth]{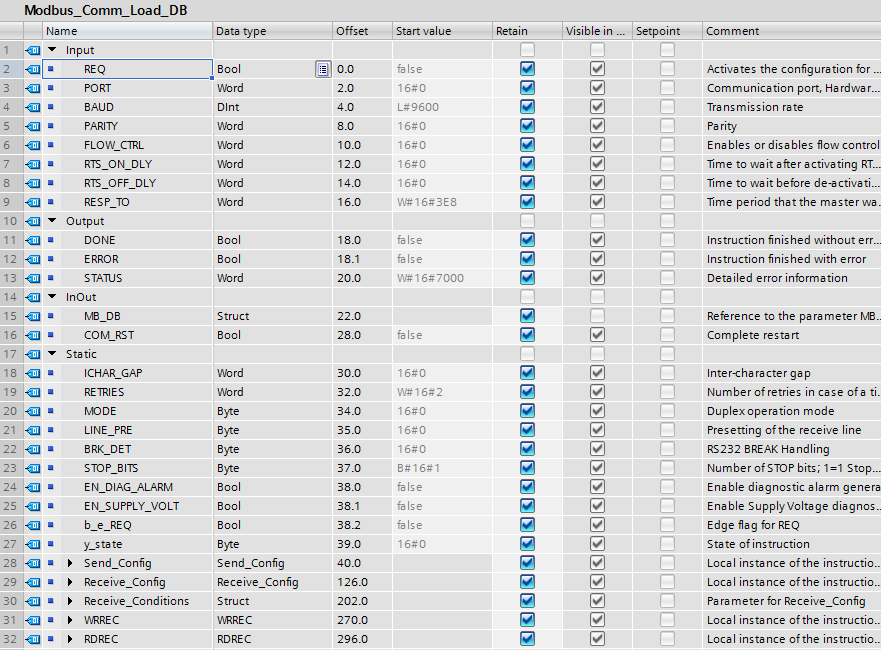}
    \vspace{-2\baselineskip}
    \caption{Modbus Library Function Data Block}
    \label{fig:modbusfunctiondb}
    \vspace{-15pt}
\end{figure}

Figure~\ref{fig:modbusfunctiondb} depicts the local DB (fVB) of library FB \textit{\textbf{Modbus\_Comm\_Load}}. This DB stores all local variables (including inputs) used by the FB. Where global variables are defined as inputs, FBs have two options. The first is to copy a global variables current state into the local DB counterpart during every control-logic cycle. The second is to configure a pointer targeting the global variables location (DB address), in this instance the pointer will be stored as a local variable within the FBs DB. The latter option is typically used where larger data inputs are required (for storage and performance efficiencies).

\vspace{-15pt}
\subsection{The Application of PCaaD}
\label{practicalSiemens}
\vspace{-5pt}
The following subsections describe how control-logic can be leveraged by attackers to achieve PCaaD via library DB/FB enumeration, leading to our three attack cases: (1) \textit{Exfiltrate Function Block Variables}, (2) \textit{Targeted Manipulation of Function Block Operations}, and (3) \textit{A Storage Based Covert Channel}. To establish communications with our PLC, the Python SNAP7 library was used~\cite{Molenaar2013}. This library allows for the crafting of Siemens S7 packets, the primary network protocol used by the PLC, and affords us with the ability to issue requests (e.g. \textit{\textbf{Read, Write, and Upload}}) as per vendor specifications~\cite{Kleinman2014}.

\vspace{-10pt}
\subsubsection{\textbf{Enumeration Phases 1 and 2}}
\vspace{-5pt}
\label{enumeration}
As described in Section~\ref{PLCbackground}, PLCs provide network access to VBs for use by HMIs, TESs, etc. For our PLC this means direct access to DBs (local and global). For example, in Figure~\ref{fig:modbusfunctionladder}, we provided the input \textit{\textbf{wRST\_OFF\_DLY}} at address \textit{\textbf{DB1.DBW0}}, this address would be used during the configuration of HMIs, allowing operators to read and depict its current state on a graphical display. 

The following three data retrieval techniques discussed in Section~\ref{PLCbackground} (\textit{Metadata, Bulk Transfer,} and \textit{Memory Address Interrogation}) exploit access granted to FB DBs over the network in order to enumerate their associated FBs, this can be considered an information leakage vulnerability. To recap, these techniques can be applied to each of the four attack vectors described in Section~\ref{Threatmodel}, where we state "execute malicious commands".

\begin{figure}[]
    \centering
    \includegraphics[scale=0.50]{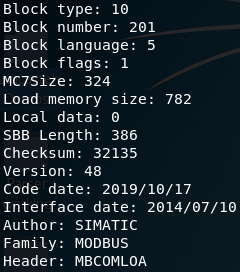}
    \vspace{-1\baselineskip}
    \caption{Get Block Info Example}
    \label{fig:getblockinfo}
    \vspace{-15pt}
\end{figure}

\textbf{Metadata (Get Block Info) -}
The first technique one can apply towards the enumeration of a FBs DB, makes use of the inbuilt PLC feature \textit{\textbf{Get Block Info}}, allowing for the extraction of metadata parameters, as can be seen in Figure~\ref{fig:getblockinfo}. The family and header fields are of greatest importance, allowing us to ascertain which FB is using the DB to store its local variables. In the example provided here, this DB is used by a \textit{\textbf{MODBUS}} family FB, so has an affiliation with Modbus communications. The header \textit{\textbf{MBCOMLOA}} is a shortened tile for the related FB \textit{\textbf{Modbus Comms Load}}, previously described in Figure~\ref{fig:modbusfunctionladder}. Therefore, it can be established that this DB is being used by the \textit{\textbf{Modbus Comms Load}} FB.

While this technique allows for the enumeration of a DBs associated FB, its reliance on the built in network function \textit{\textbf{Get Block Info}} impacts its detectability. In monitoring a \textit{\textbf{Get Block Info}} request as it traverses the network, Wireshark's~\cite{Wireshark2020} in-build protocol recognition is able to clearly identify its purpose. This is also true of nextgen security products~\cite{Checkpoint2020, claroty}. As this request is not commonly used within live industrial networks, it would raise a red flag, and could be blocked as part of an environments default security configuration profile.

\textbf{Bulk Transfer (Block Upload) -}
The second technique we have identified makes use of the inbuilt feature \textit{\textbf{Upload}}. This is a network function constructed to extract POUs in their entirety from the PLC. With PLCs, it is important to note that in certain situations we talk from the device's perspective. This is industry derived terminology consistent between vendors. Therefore, when using the term \textit{upload}, we are referring to the PLC uploading data to the user, not the user uploading data to the PLC.

\begin{figure}[h!]
    \centering
    \includegraphics[scale=0.50]{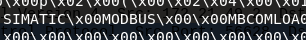}
    \vspace{-1\baselineskip}
    \caption{Block Upload Example}
    \label{fig:blockupload}
    \vspace{-10pt}
\end{figure}

In sending a DB \textit{\textbf{Upload}} request to the PLC, the entire byte-code of that DB will be returned. We examined this byte-code, and found the previously discussed family and header parameters stored in clear text (see Figure~\ref{fig:blockupload}). Running a parser over the byte-code allows us to clearly identify the DBs related FB.

As with \textit{\textbf{Get Block Info}}, the \textit{\textbf{Upload}} function has challenges aligned to detectability. While this request is more common, it only occurs when an engineer requires a copy of PLC control-logic. Therefore, where an engineer is not present, its identification on the network would raise a red flag. It too could form part of an environment's default security configuration profile, requiring an engineer to connect directly with each PLC, rather than from a remote location.

\textbf{Memory Address Interrogation (Read Requests) -}
\label{readreq}
To avoid detection challenges identified with the two former enumeration techniques, we have developed an approach based on \textit{\textbf{Read}} requests. This is a specific network function applied to the extraction of variable data from the PLC. HMIs, for example, will execute \textit{\textbf{Read}} requests to PLCs in order to extract data for use by operators. This makes it extremely common, and thus if observed on the network would be considered \textit{normal} operational behavior (i.e. \textit{normal} network traffic).

Siemens allocate static memory structures for our PLCs FBs in a minimum of 16bit blocks, even where only a single bit is required. This form of memory allocation can be observed in Figure~\ref{fig:modbusfunctiondb}, where the \textit{\textbf{REQ}} variable (boolean) resides at address 0.0 (byte 0, bit 0), and the next variable, \textit{\textbf{PORT}} (word), starts at address 2.0, i.e. byte 0 bit 1 to byte 1 bit 7 are all unused and populated with null space.

The size of some library FB DBs raises challenges in mapping all possible variable data combinations, without first conducting a tedious manual review of all possible state combinations (i.e. the more variables, the more valid variable state combinations). While machine learning approaches, on initial inspection, appear to be a feasible solution, they would rely on the ability to capture all variable state combinations. Dependent upon the FB, some variables are set once and do not change (e.g. BAUD in Figure~\ref{fig:modbusfunctiondb}). Therefore, static variables such as these would require manual updating to allow for a complete picture of all possible variable state combinations to be captured. As such, mapping all unused null spaced memory offers a viable alternative, with an increased level of performance due to the focused comparison of unused memory alone.

Through the development of a comprehensive signature set (rainbow table) based on an offline analysis of all library FB DBs (examining their structure within the Siemens TIA programming agent, as seen in example Figures~\ref{fig:modbusfunctiondb},~\ref{fig:userpassdb}, and~\ref{fig:countupdb}), focusing on their overall size and the location of any unused memory, we are now able to enumerate any Siemens library FB aligned to a DB using only \textit{\textbf{Read}} requests. The ability to achieve DB and associated FB enumeration through the use of \textit{\textbf{Read}} requests alone, offers a stealthy and effective technique when compared to the aforementioned approaches.

Our approach begins by \textit{\textbf{Reading}} every DB byte into an array. This allows us to ascertain DB size, from which we have a view of possible associated FBs based on our signature set (i.e. we have narrowed the scope of possible FBs due to their static DB size). We then check values at defined offsets (indexes within the array), where we expect to see null states (again, based on our signature set) e.g. unused offset/byte 1 as previously noted in Figure~\ref{fig:modbusfunctiondb}. Dependent upon the requirements and construct of a FB and its DB, there can be in excess of 10 complete bytes of unused memory, in addition to multiple instances of partially used bytes (i.e. up to 7 bits of unused memory within a byte).

Given current programming practices, we have demonstrated how remote control-logic enumeration can be achieved using only \textit{\textbf{Read}} requests, thus achieving stealthy PCaaD. This is significant, with Siemens providing library FBs spanning a number of critical areas (see Table~\ref{tab:examplefunctions} for a small selection of example FBs), including Communications, PID (Proportional, Integral and Derivative) Control, Safety, Remote Administration, and Alerting. The following subsections explore ways in which this information can be built upon by an attacker, to further develop their level of process comprehension, and execution of targeted attacks.

\vspace{-10pt}
\subsubsection{\textbf{Attack 1: Exfiltrate Function Block Variables}}
\vspace{-5pt}
Having successfully enumerated a local DB based on its association to a given library FB, it becomes possible to exfiltrate the data it contains. There are two distinct techniques (Direct and Pointer Decoding) that can be applied to the exfiltration of a FBs data. These techniques are based on how the FB obtains/creates data stored within its local DB. To recap, these techniques can be applied to each of the four attack vectors described in Section~\ref{Threatmodel}, where we state "execute malicious commands".

\textbf{Direct Read Requests -}
\label{directread}
Taking the Modbus Comm Load function from Figure~\ref{fig:modbusfunctionladder}, and its associated DB in Figure~\ref{fig:modbusfunctiondb}, there are a number of local variables that may be of interest to an attacker. For example, at offset 4 there exists a double integer storing the baud rate setting. Knowing that this DB is aligned to the \textit{\textbf{Modbus Comm Load}} function, and having analyzed the structure of the DB offline in TIA portal to understand its contents, we can construct a \textit{\textbf{Read}} request specifically targeted at this offset to obtain the current baud rate. 

If the variable in question is being stored in its entirety within the FBs local DB, a \textit{\textbf{Read}} request targeted directly at the variable location can be used to extract its current state. For example, to extract the baud rate variable from the Modbus Comm Load DB in Figure~\ref{fig:modbusfunctiondb}, we would construct our request to \textit{\textbf{Read}} 32 bits (double integer) from DB 1, starting at a byte offset of 4. This information develops the attackers level of process comprehension and build a picture of device-to-device data flows~\cite{Green2016}, of value when constructing a targeted attack.

\textbf{Pointer Decoding -}
\label{pointdeconstruct}
As previously described, should a function's input variable require a large block of memory, a pointer will be applied. Therefore, should we wish to exfiltrate data from a pointer input, we must first obtain the pointer address. This is achieved by constructing an initial \textit{\textbf{Read}} request targeting the pointers location within the FBs local DB. We must then decode the pointer address, prior to the formulation of an additional \textit{\textbf{Read}} request targeting the newly decoded address.

\begin{figure}[]
    \centering
    \includegraphics[width=\linewidth]{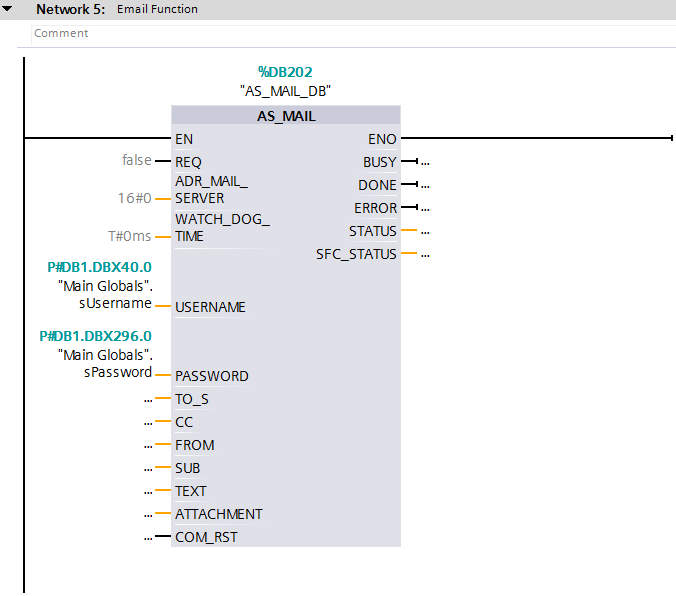}
    \vspace{-2\baselineskip}
    \caption{Email Library Function}
    \label{fig:emailfunction}
    %\vspace{-10pt}
\end{figure}

The construct of pointers within the Siemens ecosystem is unique, more specifically, they contain an address based on the Siemens addressing scheme. Taking the variable \textit{\textbf{sUsername}} from Figure~\ref{fig:emailfunction} as an example, this has an address of \textit{\textbf{DB1.DBX40.0}}, it is this address (the variables starting bit) that would be placed inside a pointer. Note the \textit{\textbf{P\#}} before the address, this denote the use of a pointer.

\begin{figure}[]
    \centering
    \includegraphics[width=\linewidth]{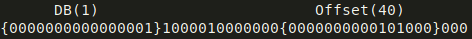}
    \vspace{-2\baselineskip}
    \caption{Pointer Structure}
    \label{fig:pointer}
    \vspace{-20pt}
\end{figure}

\begin{figure}[]
    \centering
    \includegraphics[width=\linewidth]{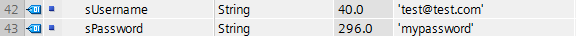}
    \vspace{-2\baselineskip}
    \caption{Username and Password Global DB}
    \label{fig:userpassdb}
    \vspace{-15pt}
\end{figure}

Figure~\ref{fig:pointer} provides an example pointer structure, highlighting the address location of \textit{\textbf{DB 1}} with an offset of \textit{\textbf{40}}. With this information we can begin constructing a new \textit{\textbf{Read}} request. To complete our request, we must also ascertain the target variable size. An offline analysis of the FB input will provide this information. For example, Figure~\ref{fig:emailfunction} is a library FB taking email username and password details, these must be strings. Strings have a standardized size of 256 bytes, as can be see in Figure~\ref{fig:userpassdb} (the location of this pointer). Therefore, \textit{\textbf{Reading}} 256 bytes from \textit{\textbf{DB 1}} at a byte offset of \textit{\textbf{40}}, will provide us with the username \textit{\textbf{test@test.com}}. Likewise, \textit{\textbf{Reading}} 256 bytes from \textit{\textbf{DB 1}} at a byte offset of \textit{\textbf{296}}, will provide us with the password \textit{\textbf{mypassword}}. Both the username and password are stored in clear text. The use of this information to an attacker could prove highly valuable, especially where the exfiltrated credential set is applicable across multiple systems within the target environment.

The two exfiltration techniques described here cover all variables stored within a FBs local DB. While the exfiltration of baud rate and credentials demonstrate a clear benefit to attackers, they represent just two examples from a much larger set (i.e. thousands) of FB variables across the Siemens library. With Siemens library FBs spanning an array of capabilities as noted in Section~\ref{enumeration}, and examples provided in Table~\ref{tab:examplefunctions}, the wider ramifications of data exfiltration, particularly with regards to the development of process comprehension, is significant.

\vspace{-10pt}
\subsubsection{\textbf{Attack 2: Targeted Manipulation of FB Operations}}
\label{manfunbehaviour}
\vspace{-5pt}
The use of \textit{\textbf{Write}} requests are required to manipulate FB behavior through the targeting of local DB variables. These requests are typically seen where operators modify operational process behavior via a HMI (e.g. start/stop a pump). They are, therefore, common permissable commands on an industrial network. To recap, this technique can be applied to each of the four attack vectors described in Section~\ref{Threatmodel}, where we state "execute malicious commands".

While \textit{\textbf{Read}} requests have no limitations in their ability to execute as expected, the ability to successfully manipulate FB behavior has one: cycle time. Where variable states are updated during every control-logic cycle, overwriting them with a 100\% success rate becomes a challenge~\cite{Robles-Durazno2019}. This is the case for Siemens FBs inputs, with input states moved into a FBs local DB during every cycle. Using Figure~\ref{fig:modbusfunctionladder} as an example, Section~\ref{SiemensEco} discussed the following three techniques to provide FB inputs: directly, from a global DB, and through the use of default values. Should a PLC programmer apply the first or second technique, input states will be written to the FBs local DB during every cycle of control-logic. Where default values are used, this limitation does not exist, and a singe \textit{\textbf{Write}} request will overwrite their state.

\begin{figure}[h!]
    \centering
    \includegraphics[width=\linewidth]{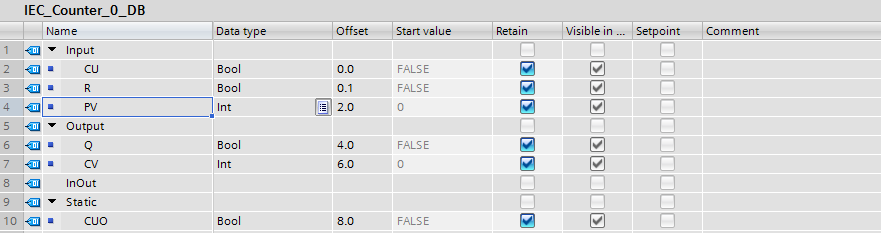}
    \vspace{-2\baselineskip}
    \caption{IEC Count Up DB}
    \label{fig:countupdb}
    \vspace{-10pt}
\end{figure}

The cycle time limitation can be circumvented under certain conditions. For example, the \textit{\textbf{IEC\_CU}} FB (see Figure~\ref{fig:countupdb} for this FBs local DB structure) takes a boolean input (byte 0, bit 0) as a trigger, and provides an integer count output (byte offset 6). Upon a state change of this boolean trigger, the integer count output increases by one. There is a second input responsible for resetting the current count value back to 0 (byte 0, bit 1). Should a PLC programmer allocate a global variable to this input, an attacker would not be confidently able to target the local DB address with a single \textit{\textbf{Write}} request to reset the current count. However, variables used within the FB and the FBs outputs can provide an alternative target. Writing a single \textit{\textbf{0}} to the current count integer (byte offset 6), would reset the count without any limitations. This is a vulnerability induced through the way in which the FB code has been written, something only the vendor can address.

The ability to execute \textit{\textbf{Write}} requests to a FBs local DB, opens the door to wide-ranging impact. Using the \textit{\textbf{Modbus\_Comm\_Load}} function discussed in Section~\ref{directread} as an additional reference point, an attacker could target the BAUD variable, placing the Modbus communication channel into a defective state. Dependent upon what the channel is being used for, this could impact the PLCs ability to communicate with other PLCs for critical operational data exchanges, or even prevent operators from receiving alarms. As with data exfiltration, the quantity and functionality offered within the Siemens FB library brings with it a significant operational risk due to the described manipulation (e.g. the manipulation of Safety Functions, of paramount importance in protecting human life).

\vspace{-10pt}
\subsubsection{\textbf{Attack 3: A Storage Based Covert Channel}}
\label{practicalc2}
\vspace{-5pt}
In developing our enumeration technique based on \textit{\textbf{Read}} requests, we focused on the identification of local FB DBs through unused memory. In the previous section on manipulating FB operations, we have demonstrated an attackers ability to execute \textit{\textbf{Write}} requests targeting a FBs local DB. The remainder of this section takes attack vectors 3 and 4 from Section~\ref{Threatmodel}, where security zoning has been established, and presents a method by which it can be violated through the combination of these two concepts.

\textbf{Scenario -}
Unlike Attacks 1 and 2, that require the attacker to obtain access to a PLC alone, a broader set of pre-requisites are required for this attack. As described in attack vectors 3 and 4 (See Section~\ref{Threatmodel}), the Field Site PLC/RTU is permitted to communicate with all other Field Site devices and also the SCADA Server. All other Field Site devices are not permitter to communicate outside of the Field Site network (managed by perimeter firewalls, a baseline recognized practice to defend zone boundaries~\cite{Stouffer2015,6244333}).

Should an attacker compromise the SCADA Server (attack vector 4), allowing for direct interaction with it, and the devices it connects to, only one Field Site device would be accessible, the PLC/RTU. In contrast, should an attacker compromise the HMI (attack vector 3), all Field Site devices would be accessible. However, as the HMI is isolated within the Field Site network zone, remote interaction would not be possible. The ideal attack vector, would involve a method by which the functionality of both the SCADA Server (remote connectivity into the Field Site Network network) and HMI (access to all Field Site devices) could be leveraged. This is where their common resource, the RTU/PLC comes in, providing an ideal pivot point between the two devices. For additional clarity, the compromised SCADA Server acts as the C2-Server, and the HMI acts as the C2-Client.

\textbf{Channel Operations -}
\label{channelops}
In order to establish a covert C2 channel between the C2-Client and the C2-Server via a FBs unused memory, each must first enumerate all library FB DBs using the \textit{Memory Address Interrogation} technique defined in Section~\ref{readreq}. This technique should be adopted as it stealthy in nature, thus avoids raising an alert through alternate approaches (i.e. \textit{Metadata} or \textit{Bulk Transfer}). Once enumerated, both parties must select the same DB and unused memory offsets to begin communicating. Our approach to the selection of a DB is defined by the quantity of available unused bytes of memory i.e., both parities will use the DB that contains the most unused bytes of memory. Where two or more DBs meet this requirement with an equal quantity of unused bytes, the first DB will be selected (e.g. should DB3, DB5, and DB6 all contain 10 unused bytes, DB3 would be selected). Once selected, the first two unused bytes of that DB will be aligned to synchronization and data exchange, respectively.

Table~\ref{tab:sync} provides an overview of the binary \textbf{synchronization states} written to, and read from, the first unused byte of memory. Each party must adhere to this scheme in the establishment and continuation of communication exchanges. To reiterate, in our scenario the HMI adopts the role of \textit{\textbf{C2-Client}}, and the SCADA Server adopts the role of \textit{\textbf{CS-Server}}. 

\begin{table}[]
   \begin{center}
    \begin{tabular}{|l|l|l|}
    \hline
    \textbf{Function} & \textbf{C2-Server} & \textbf{C2-Client} \\ \hline
    Hello             &                 & 00000001        \\ \hline
    Hello Ack         & 00000011        & 00000000        \\ \hline
    Write             & 01000000        & 11100000        \\ \hline
    Reading           & 11110000        & 01100000         \\ \hline
    Read              & 00000000        & 00000000        \\ \hline
    Final Write       & 11111111        & 11111110        \\ \hline
    On Hold           & 00011000        & 00011000        \\ \hline
    \end{tabular}
    \end{center}
    \vspace{-17pt}
    \caption{Synchronization Byte}
    \label{tab:sync}
    \vspace{-15pt}
\end{table}

A "what goes in goes out" approach has been applied in our current covert C2 channel. The C2-Server's operator (the attacker) will construct a terminal request it wishes the C2-Client to execute (e.g. ping 192.168.0.1). This request will be cut into individual characters, each of which are sent over the data exchange byte. To achieve this, each character is converted into its decimal counterpart (ASCII character encoding). Each character is reconstructed by the C2-Client, and once the entire request has been received, it is executed in the C2-Client's terminal. The terminal's response is then cut into individual characters, sent back to the CS-Server, reconstructed, and displayed.

While the design of this covert C2 channel data exchange appears simplistic, it is effective, and acts as a demonstrable tool in the use of unused FB memory as a pivot point to violate security zoning. Its application across Siemens library FB DBs is widely applicable, however, in order to ensure it does not impact a FBs operation, a series of tests were undertaken. Through this testing, we identified that in some instances where only one bit within a two byte block is allocated, the FB would look for state changes across all sixteen bits. Therefore, the use of unused memory would have an undesirable impact on FB operation, creating issues with the stable and expected execution of control-logic, and increase the chance of detectability. Fortunately, we have found few instances that impact ones ability to establish a covert C2 channel through the use of this technique. Furthermore, in monitoring the status of the primary bit in use, if it is set to \textit{\textbf{1}}, the use of unused memory will have no impact on FB operation.

\vspace{-10pt}
\subsection{Cross Vendor Generalization}
\vspace{-5pt}
The PoC running example described throughout this section has been aligned to the Siemens 300 series PLCs. Considering alternative Siemens PLC series, 400 series PLCs act as a mirror to the 300 series, making the techniques described here holistically applicable. ET200 series PLCs act as a bridge between the 300/400 series and the newer 1200/1500 series. Testing with an ET200S resulted in the direct applicability of our techniques. Finally, for 1200 and 1500 series PLCs, where library functions (e.g. \textit{\textbf{TMAIL\_C}}) apply direct addressing by default, the described techniques are also applicable.

In the consideration of PLCs from other vendors, we conducted some initial experimentation, and although the ecosystem of the devices tested differs somewhat to Siemens, our fundamental concept (see Section~\ref{PLCbackground}) holds true to ascertain similar attacks. In addition, we identified vendors providing library functions for one another's devices. ABB, for example, provide library functions supporting drive integration with Siemens PLCs. These library functions allow a Siemens PLC to control an ABB drive~\cite{ABB}. The FBs and associated DBs provided by ABB, harbour the same deficiencies as discussed in Section~\ref{practicalSiemens}. Therefore, ABB devices become exploitable through this integration.

The previous subsections have described how current PLC programming practices can be exploited by attackers to achieve PCaaD and develop targeted attacks. Furthermore, each technique can be fully automated, and is applicable to any environment (e.g. water, energy, a testbed). The following section provides a broader discussion around the defined problem space, including mitigation techniques, a process for fully automated system exploitation, and vendor response.

\vspace{-10pt}
\section{Lessons Learnt}
\label{discussion}
\vspace{-10pt}
We now overview salient points on PLC exploitability via the PCaaD process, highlight issues in PLC programming practices, offer potential mitigations, provide a response from Siemens, and detail a process for automation PCaaD and attack execution.

\vspace{-10pt}
\subsection{The Impact of PCaaD}
\vspace{-5pt}
This work demonstrates the theoretical and practical application of PCaaD, targeting only the PLC. Existing approaches to develop a high level of process comprehension can be lengthy and involve data aggregated from multiple sources~\cite{Green2017a}. This aggregation of data is largely applied toward an attackers understanding of PLC control-logic. With a high level of process comprehension, targeted operational process manipulation is made possible. Without it, attackers are limited to primitive DoS attacks. The capabilities described in this paper provide increased process comprehension, and therefore the ability to strategically attack a PLC. While this does not provide full process comprehension, it demonstrates an approach which supports enhanced attack complexity (e.g. \textit{data exfiltration, targeted manipulation of FB variables,} and \textit{covert channel creation}). The described techniques can be adopted in parallel to existing approaches~\cite{Green2017a}, enhancing an attackers understanding of PLC functionality.

This can be exemplified when comparing targeted manipulation of FB operation attacks, to attacks demonstrated in previous work~\cite{Robles-Durazno2019}, reducing the requirement for priori knowledge of a system to identify target variables, and also the need for rapid remote overwriting of PLC memory locations to maintain the desired effect (i.e. the cycle time limitation).

In addition to the increased level of enumeration/attack sophistication available through the use of PCaaD, this paper practically demonstrates the use of a PLC in a covert channel. It was through the identification of null space signatures during the PCaaD process, that gave rise to the possibility of a storage based covert channel. As such, the PLC can be used as a pivot point between protected Field Site networks and external networks. The exploitation of a PLC in facilitating an attack, rather than being the target of an attack, along with the mechanics of this storage based covert channel, are both novel. Importantly, this adds a previously seldom considered class of security challenge for PLC implementation, suggesting a reconsideration in the understanding of ICS security zoning, a primary defensive measure~\cite{6244333}. Applying the described approach highlights challenges in conventional segregation techniques to adequately prevent attackers from accessing "isolated" zones. This work practically demonstrates channel feasibility, and as such forces the ICS community to think differently about the role of PLCs in cyber attacks. 

\vspace{-10pt}
\subsection{Mitigations}
\label{mitigation}
\vspace{-5pt}
This work continues to open the door of a new vulnerability class, one derived through the creation of insecure control-logic. Using popular FB libraries as an example, we have been able to articulate a set of independent and unique attacks, which have been collectively applied to support an end-to-end autonomous exploitation process flow (see Section ~\ref{PCaaDProcess}). The exploitability of FB libraries is, in part, tied to their static and accessible memory structures, but also in how they have been written by vendors, and implemented by PLC programmers. Furthermore, should PLC programmers develop their own custom FBs (similar to those provided by ABB), they too could offer a target for attackers through the adoption of our approach. 

During the development of our PCaaD approach, and each exploitation technique, mitigations were actively sought. Here a number of approaches are presented that would aid in the reduction of PCaaD and associated exploitation susceptibility. Some examples are aligned to the Siemens eco system to support discussion, but provide a valid viewpoint for other vendors.

{\textbf{Network Access Restrictions}} form a well explored starting point towards appropriate mitigation~\cite{Stouffer2015, 6244333}. If network-level access to fVBs can not be natively restricted, ensuring PLCs are isolated from widely accessible networks forms a baseline requirement. In addition to this, deploying next generation IDS/IPS with industrial protocol recognition~\cite{Checkpoint2020, claroty}, allows for granular access control between trusted connections. Limiting fVB access could therefore be applied within a baseline rule-set.

{\textbf{Stack Canary}} usage within conventional applications is common practice. Acting as a flag for the identification of unexpected data within a defined memory structure. A similar concept could be applied as part of a PLCs control-logic, through the creation of a custom FB to monitor critical data points within library fVBs. For example, in monitoring the BAUD variable in Figures~\ref{fig:modbusfunctionladder} and~\ref{fig:modbusfunctiondb}, should its state remain static at 9600, a single rung of validation logic would be required. This rung would check the current value during every cycle, if the value did not equal 9600, it would set a bit, thus raising an alert on local HMIs. This concept could also be applied to all null spaced memory, raising an alert if it becomes populated with data.

{\textbf{Setting All FB Inputs}} even where default values are applicable, will induce the cycle time limitation discussed in section~\ref{manfunbehaviour}. While this will not mitigate all attacks, it presents an additional challenge for attackers to overcome.

{\textbf{Filling Unused Memory}} with random data would make enumeration via read requests more challenging.

{\textbf{Additional Checks for Neighboring PLCs}} could be applied to validate received data. For example, considering the count function attack in section~\ref{manfunbehaviour}, an additional check to identify state changes on the reset bit, before accepting a reset of the count value, would provide an additional validation.

{\textbf{Read/Write Protection}} features within 300, 400, and ET200 series devices would prevent enumeration using the \textbf{\textit{Upload}} technique. However, the applicability of alternate options discussed in section~\ref{enumeration} would remain unaffected. 

{\textbf{Vendor Centric}} changes could be applied to the development and inclusion of library FBs. For example, FBs could be written in such a way as to allow for all data generated to be made available as defined outputs, fed into gVBs, meaning no local or remote access to a FBs fVB would be required. In addition, Siemens implement know-how protection on FB code, preventing programmers for viewing pre-compiled logic. This principle could be applied to a FBs associated fVB, making it harder to map memory usage and generate signature sets.

Memory allocation within fVBs could be constructed in a more efficient way, reducing unused capacity, and thus advancing the technical requirements for successful enumeration using read requests alone, and limiting attackers ability to create covert channels. 

A blanket rejection of all network-based requests targeting fVBs would offer holistic mitigation. This feature can be manually enabled (disabled by default) on gVBs used by 1200 and 1500 series devices, but is not yet available for fVBs.

{\textbf{Enable Block Optimization}} on Siemens 1200 and 1500 series PLC where possible (not currently available across all library FBs). This feature removes static addressing schemes, with variables referenced based on name as opposed to address. It may still be possible to enumerate and attack these functions based on the use of standardized variable names, however this is outside of our current scope. The only drawback of this feature is that third-party devices (e.g. HMIs) requiring access to PLC data must support S7 Comm Plus. Where support is not available, a blanket disabling of block optimization may be adopted by PLC programmers, thus re-introducing static addressing schemes.

\vspace{-10pt}
\subsection{Vendor Response}
\vspace{-5pt}
Following responsible disclosure practices, we contacted Siemens and ABB to make them aware of the issues identified in this paper. Siemens response was as follows:

"The flat addressing in PLCs like S7-300 and ET200S CPU is a design decision from the 90s and cannot be easily changed without breaking existing installations. Siemens recommends customers to restrict network access to the affected devices, to apply Defense-in-Depth measures that can be found in the Operational Guidelines for Industrial Security, and to follow the recommendations in the product manuals. Siemens improved this behavior in the new PLC generation (S7-1200 and S7-1500) by creating the optimized Data Blocks and additional levels of protection to these PLCs."

It is worth noting, Siemens have committed to offering 300 series devices until 2023, with an additional ten years of support beyond this point~\cite{Siemens2020a}.

\vspace{-10pt}
\subsection{Automated Evaluation of Control-Logic}
\label{PCaaDProcess}
\begin{figure}[t!]
	\centering
	\includegraphics[width=6cm]{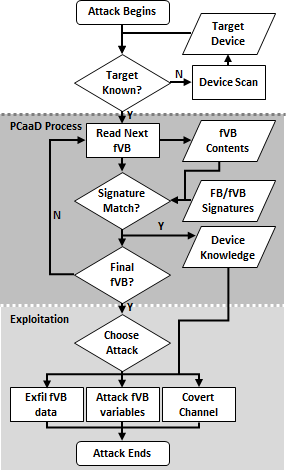}
	\vspace{-1\baselineskip}
	\caption{Automated PCaaD \& Attack Execution Process}
	\label{fig:tool}
	\vspace{-15pt}
\end{figure}

\vspace{-5pt}
The PCaaD capabilities presented here represent a first step towards the ability to automate targeted exploitation of operational processes and PLC configuration parameters. The endpoint of this research is to provide an integrated exploitation platform. Such a platform would enumerate using the identified, and expanded upon PCaaD approaches, and further integrate exploitation components extending beyond those already identified. Embedding this functionality into a single platform forms a linear attack offering, heightening each component/techniques collective value. This has been seen in IT penetration testing (e.g. Metaspolit). Figure~\ref{fig:tool} provides a high-level process flow, depicting the functionality offered through the integration of each component described within in this paper. A PoC tool operating as outlined in Figure~\ref{fig:tool} has been developed in Python by the research team.

The process incorporates manual and automated target selection (the latter is achieved through use of PLC Scan~\cite{Efanov2017}). Once a suitable target PLC is identified, the PCaaD process is initiated to enumerate control-logic and identify possible target memory locations for our three exploit categories. Additional flexibility is included to add signatures beyond the known vendor provided library FBs. Custom, in-house signatures can be added to the repository, supporting the enumeration and exploitation of in-house developed FBs/fVBs. As a whole, this platform will aid an organizations efforts to better understand the scale of exploitable control-logic within their estate, and to evaluate security zoning.

\vspace{-10pt}
\section{Conclusion and Future Work}
\vspace{-10pt}
\label{conclusion}
This work has demonstrated the feasibility of stealthy, sophisticated, targeted attacks on industrial processes with no prior knowledge of the target PLCs configuration or control-logic. An attack of this nature was previously considered to be impractical. However, this paper demonstrates that through the exploitation of current PLC programming practices, and code reuse patterns, such attacks are possible. As a result, several security challenges are presented. These challenges align to successful PCaaD giving rise to sophisticated targeted attacks against previously unseen industrial processes, and the use of a PLC in the facilitation of attacks via a storage based covert channel.

A further benefit of the PCaaD approach described in this paper is the ability to fingerprint custom FBs. These FBs, written in-house by PLC engineers for deployment across an organizations operational estate, can now be identified using the our signature techniques. Given the wide-spread use of identical custom FB libraries within an organization~\cite{Ljungkrantz2007}, identifying custom control-logic offers added value, increasing the breadth of FB detection beyond publicly accessible/vendor provided libraries.

While our practical proof of concept focused on demonstrating the identified security challenges on Siemens PLCs, an initial exploration of two other prominent vendors highlights key parallels. The functional similarities between vendors, suggests other vendors' devices to be equally exploitable.

A selection of appropriate mitigation techniques have been discussed. This included points raised by Siemens, aligned to capability embedded within their latest product range. However, we believe features within these products could be circumvented, offering an additional direction for future work.

The research agenda for future work in this space will focus on two primary themes: Improvements in PCaaD techniques, and exploring further possibilities to utilise PLCs as an attack platform.

For PCaaD techniques, our first phase of work will be to widen the empirical exploration to a broader range of vendors. The aim of which is to develop a comprehensive tool, able to target and enumerate fVBs across a range of PLCs. To achieve this, further developments in identifying memory features to provide more sophisticated signatures may be required. This includes the exploration of machine learning approaches.

The use of PLCs as a tool during an attack will be further developed. This will focus on establishing a range of mechanisms to enhance our existing covert channel. More specifically, we will explore the trade off between channel features, detectability, robustness, and control overhead.

%-------------------------------------------------------------------------------
\bibliographystyle{plain}
\bibliography{refer}
\onecolumn
\section*{Appendix}
\label{Appendix}
\vspace{-15pt}

\begin{table}[H]
    \begin{tabularx}{\textwidth}{|l|l|l|}
    \hline
    \textbf{FB Category} & \textbf{Example FB} & \textbf{Description}                                                                                                                                                                        \\ \hline

    Basic Instructions         & TON                       & Delays the setting of the Q parameter for the programmed duration PT                                                                                                                      \\ \cline{2-3}
                               & CTU                       & Increments the value at the CV parameter                                                                                                                                                   \\ \cline{2-3}
                               & SMC                       & Compares the signal state of up to 16 programmed input bits \\ && (IN\_BIT0 to IN\_BIT15) with the corresponding bits  of the comparison \\ && masks for each step                                   \\ \cline{2-3}
                               & ACK\_GL (Safety)          & Creates an acknowledgment for   the simultaneous reintegration of all \\ && F-I/O or channels of the F-I/O of an F-runtime group after communication \\ && errors, F-I/O errors, or channel faults \\ \cline{2-3}
                               & SFDOOR (Safety)           & Safety door monitoring                                                                                                                                                                    \\ \hline

    Extended Instructions      & SET\_SW                   & Supports the switch from daylight-saving time to standard time in CPUs \\ && that are not equipped with time-of-day status                                                                   \\ \cline{2-3}
                               & TIMESTMP                  & Transmits messages with a time   stamp of an IM153-2 to its instance DB                                                                                                                    \\ \cline{2-3}
                               & RDREC                     & Reads the data record with the   number INDEX from the component \\ && addressed using the ID                                                                                                   \\ \cline{2-3}
                               & SETIO                     & Consistently transfers the data   from the source area spanned by \\ && OUTPUTS to the addressed DP standard PROFINET IO device, and, if \\ && necessary, to the process image               \\ \cline{2-3}
                               & PACK                      & Transfers data located between   any addresses and a table                                                                                                                              \\ \hline

    Technology                 & CONT\_C (PID)                  & Controls technical processes with continuous input and output variables                                                                                                                 \\ \cline{2-3}
                               & PULSEGEN (PID)                 & Implements a fixed setpoint controller with a switching output for \\ && proportional actuators                                                                                               \\ \cline{2-3}
                               & MC\_MoveAbsolute          & Starts an axis positioning motion to move it to an absolute position                                                                                                                   \\ \cline{2-3}
                               & EncoderSINAMICS           & Integrates a SINAMICS drive in Easy Motion Control                                                                                                                                       \\ \cline{2-3}
                               & OVERRIDE (PID)                 & Implements an override control                                                                                                                                                          \\ \hline

    Communication              & PUT                       & Writes data to a remote CPU if the connection does not take place via a CP                                                                                                             \\ \cline{2-3}
                               & \begin{tabular}[c]{@{}l@{}}PG\_DIAL (Remote \\ Administation)\end{tabular} & \begin{tabular}[c]{@{}l@{}}Transfers a telephone number and an event ID to a TS Adapter. Using the\\ specified telephone number, the TS Adapter establishes a remote\\ connection to a programming device/PC\end{tabular}  \\ \cline{2-3}
                               & MODBUSPN                  & Enables communication between a CPU with integrated PN interface and a \\ && partner which supports the Modbus/TCP protocol                                                              \\ \cline{2-3}
                               & \begin{tabular}[c]{@{}l@{}}AS\_MAIL\\ (Alerting)\end{tabular}              & \begin{tabular}[c]{@{}l@{}}Uses the Simple Mail Transfer Protocol (SMTP) to transfer an e-mail from\\ a CPU to a mail server\end{tabular} \\ \cline{2-3}
                               &  \begin{tabular}[c]{@{}l@{}}SMS\_SEND\\ (Alerting)\end{tabular}             & \begin{tabular}[c]{@{}l@{}}Transfers a telephone number, a service center number and an SMS \\ message to a TS Adapter\end{tabular} \\ \hline
    \end{tabularx}
    \vspace{-10pt}
    \caption{Example Library Functions (Siemens TIA v13)~\cite{Siemens2020}}
    \label{tab:examplefunctions}
\end{table}
%%%%%%%%%%%%%%%%%%%%%%%%%%%%%%%%%%%%%%%%%%%%%%%%%%%%%%%%%%%%%%%%%%%%%%%%%%%%%%%%
\end{document}